\documentclass[aps,prl,pifont,citesort,epsf,twocolumn]{revtex4}

\usepackage{graphicx,amsmath,amssymb}
\usepackage{dcolumn}
\usepackage{bm}

\newcommand{\bk}[1]{\left ( #1\right )}
\newcommand{\eq}[1]{\begin{equation} \newline #1 \end{equation}}
\newcommand{\eqn}[1]{\begin{eqnarray} \newline #1 \end{eqnarray}}

\newcommand{\hs}{\hspace{0.2cm}}

\newcommand{\bra}[1]{\langle#1|}
\newcommand{\ket}[1]{\left |#1\right \rangle}

\newcommand{\var}[1]{\left < #1 \right >}

\newcommand{\nn}{\nonumber}
\newcommand{\half}{\frac{1}{2}}
\newcommand{\adag}{\hat{a}^{\dag}}
\newcommand{\anih}{\hat{a}}
\newcommand{\mat}[4]{
\left[\begin{array}{cc}
#1 & #2 \\
#3 & #4
\end{array}\right]}

\begin{document}

\title{Security of Post-selection based Continuous Variable Quantum Key Distribution against Arbitrary Attacks}

\author{Nathan Walk}
\email{walk@physics.uq.edu.au}
\author{Timothy C. Ralph}
\affiliation{
Centre for Quantum Computation and Communication Technology\\
 School of Mathematics and Physics, University of Queensland, St Lucia, Queensland 4072, Australia}
\author{Thomas Symul}
\author{Ping Koy Lam}
\affiliation{Centre for Quantum Computation and Communication Technology\\
Department of Physics, Faculty of Science, Australian National University, ACT 0200, Australia}

\date{\today}

\begin{abstract}
We extend the security proof for continuous variable quantum key distribution protocols using post selection to account for arbitrary eavesdropping attacks by employing the concept of an equivalent protocol where the post-selection is implemented as a projective quantum measurement.  We demonstrate that the security can be calculated using only experimentally accessible quantities and finally explicitly evaluate the performance for the case of a noisy Gaussian channel in the limit of unbounded key length.
\end{abstract}

\maketitle

Quantum key distribution (QKD) is the process of generating a common random key between two parties using a quantum communications protocol.  The power of this method is that the security of the key distribution, and the subsequent communication via a one time pad, is established while making no assumptions about the technological capabilities of a eavesdropper.  This procedure also has the distinction of being the most developed \cite{Scarani:2009p378} quantum information technology.

There are two main flavours of QKD, discrete variable (DV) and continuous variable (CV) which are realised by encoding and then detecting single photons \cite{Zhao:2008p527} and the quadrature variables \cite{Cerf:2007p524} of the optical field respectively.  The latter kind, which we shall consider here, has the advantage of higher raw bit rates due to the high efficiency and high bandwidth of homodyne detection and ease of integration with existing communications infrastructure.  CV protocols that employ post-selection \cite{Silberhorn:2002p149} - a classical filtering of the measurement results -  enjoy additional advantages in terms of versatility and reconciliation efficiency.

Asymptotic (in the sense of string length) unconditional security for protocols that do not employ post-selection is achieved by first noting the equivalence of an experimentally implemented prepare and measure (P\&M) scheme to an entanglement based (EB) version \cite{Grosshans:2003p526}, followed by the result that for collective attacks security may be bounded from below by assuming the entangled state at the end of the protocol is Gaussian \cite{GarciaPatron:2006p381} and finally a proof that that collective attacks are asymptotically optimal \cite{Renner:2009p1}.  However for protocols using post-selection (PS) this analysis cannot be straightforwardly applied as an equivalent entanglement based picture has yet to be constructed, with security only shown under the assumption of a Gaussian eavesdropping attack \cite{Heid:2007p375}.  By filling in this gap we are able to provide a security proof for coherent state post-selection protocols.

{\it Security of CVQKD.---}In general one equates each protocol in which: the sender (Alice) prepares an ensemble of quantum states based upon a classical random probability distribution and sends it through the domain of the eavesdropper (Eve) to the recipient (Bob), to an entanglement based scheme in which: Alice prepares an entangled state one half of which is kept and used for a projective measurement; and the other is transmitted to Bob again through Eves domain. The proper choice of the initial entangled state and the projective measurement by Alice allows us to rigorously express any prepare and measure scheme  \cite{Grosshans:2003p526}.  Bob makes a quadrature measurement upon his received states and then Alice and Bob engage in a reconciliation procedure to correct the errors in their shared classical string. The secret key rate for the entire protocol is then given by 
\eq{K = \beta I(a:b) - I(E:X), \hs X \in \{A,B\}}
where $I(a:b)$ is the Shannon mutual information between classical strings belonging to Alice and Bob at the end of the protocol, $\beta$ is the efficiency of their reconciliation procedure and $I(E:X)$ is the quantum mutual information between either Eve and Bob if considering reverse reconciliation protocols or Eve and Alice if considering direct reconciliation.  Although employing a Gaussian encoding would in principle be optimal for Alice and Bob, in practice the reconciliation efficiency of such a scheme scales poorly with loss.  An alternative is to use a simpler sign encoding for which the mutual information is
\eq{I(A:B) = 1 + p_e\log_2(p_e) + (1-p_e)\log_2(1-p_e) }
where $p_e$ is the probability of Bob measuring a positive value given a negative encoding and vice versa.

Eve's mutual information is given by the difference between von Neumann entropies of the entangled state before and after Alice or Bob's measurement.  The direct reconciliation expression is {GarciaPatron:2006p381}
\eq{\label{IE}I(E:A) = S(E) - S(E|a) = S(AB) - S(B|a)}
with the von Neumann entropy given by $S(\rho) = \mathrm{tr}(\rho \log\rho)$ and for the second equality we have used the fact that overall tripartite state $\ket{ABE}$ is pure.  This quantity is not easy to calculate in general but it has been shown \cite{GarciaPatron:2006p381} that we may bound the expression from below by analysing a Gaussian state with the same first and second moments.  For Gaussian states, the von Neumann entropy is obtained straightforwardly and thus the security of the entire protocol can be characterised entirely by the covariance matrix of the entangled state shared by Alice and Bob.

{\it Equivalent post-selection scheme.---}While reverse reconciliation can be shown to to be secure for large losses direct reconciliation is only successful when the channel loss is below 50$\%$ or 3dB.  This can be remedied using post-selection \cite{Silberhorn:2002p149}, a technique in which a region is identified in the space of Bob's possible quadrature measurement results and Alice's quadrature encoding and only instances within the region are kept to form the key.  This improved performance comes at the price of not being able to directly apply Eq.\ref{IE} as this would not allow for Eve's knowledge of Alice and Bob's post-selection.  This can be accounted for as long as one can keep track of the way post-selection by one party influences the state of the other in the equivalent EB scheme which we shall now demonstrate.  

\begin{figure}[htbp]
\begin{center}
\includegraphics[width = 7.8cm]{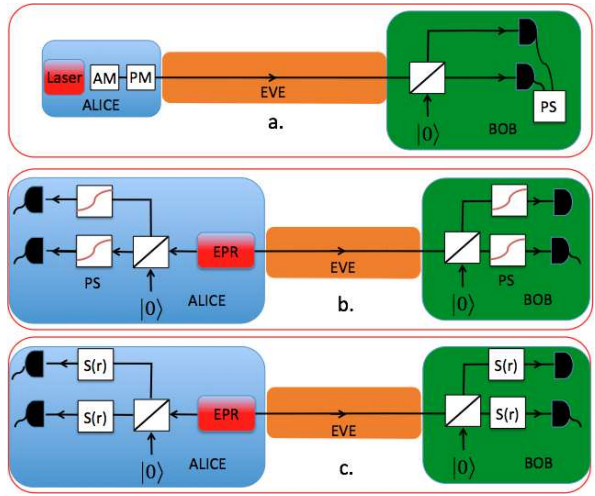}
\caption{Prepare and measure and entanglement based versions of a post-selected protocol. In the $P\&M$ scheme a), Alice sends a ensemble of coherent states through Eve's domain to Bob who heterodyne detects and applies classical post-processing. Equivalently, b), Alice and Bob pass their respective arms of a distributed EPR pair through devices that project into quadrature regions corresponding to data kept in a).  Security is calculated for an equivalent all Gaussian scheme, c), that has the same output statistics from heterodyne measurement, providing a lower bound on the secret key rate.} 
\label{schematic}
\end{center}
\end{figure}

For the sake of concreteness we shall consider a particular P\&M scheme, Fig.1 panel a), in which Alice draws values ($x_A,p_A$) from a bivariate Gaussian of 0 mean and variance $V_A$ and uses these numbers to modulate the vacuum to create an ensemble of coherent states of the form $\ket{\bar{x} + i\bar{p}}$ which she sends to Bob through a quantum channel. Bob uses heterodyne detection on his received states, measuring quadratures given by $\hat{x} = \anih + \adag$ and $\hat{p} = i(\adag - \anih)$ where we have normalised the vacuum noise to unity. Alice then filters her results such that she only keeps values whose magnitude lies between upper $U_A$ and lower $L_A$ thresholds, i.e. $L_A \leq |x_A|,|p_A| \leq U_A$ and Bob similarly post-selects his measurement results such that  $L_B \leq |x_A|,|p_A| \leq U_B$.  Finally Alice and Bob publicly announce a subset of their data to characterise the channel and, if secure, engage in reconciliation and privacy amplification to distill a completely secure key.
This is very similar to the situation that was implemented in  \cite{Lance:2005p376,Symul:2007p151} except for the fact that aside from channel characterisation, neither Alice or Bob announce anything about their measurements or encoding on a shot by shot basis and Bob employs heterodyne detection . 

The equivalent entanglement based scheme, Fig.1 panel b), involves Alice preparing a two-mode squeezed vacuum or EPR state, one mode of which she keeps and measures, the other is transmitted to Bob over a quantum channel.  At the measurement step Alice and Bob first split up their modes on a balanced beamsplitter introducing a unit of vacuum noise.  We implement the post-selection as a device that non-deterministically performs one of the projective measurements $\int_{-U}^{-L} dx \ket{x}\bra{x} + \int^{L}_{U} dx \ket{x}\bra{x}$ or $\int_{-U}^{-L} dp \ket{p}\bra{p} + \int^{L}_{U} dp \ket{p}\bra{p}$.  If the measurement fails the attempt is abandoned.  If it succeeds the state is retained and subsequently measured.  

Finally we bound the keyrate from below by evaluating the security of the third protocol, Fig.1 panel c), which has exactly the same first and second moments as the previous entangled scheme but is entirely Gaussian, allowing for the exact calculation of entropies while maximising Eve's information.  The squeezers are present to reflect the fact that Alice's post-selected classical distribution in the $P\&M$ scheme can have variance less than shot noise, thus the corresponding Gaussian protocol must have the potential to be squeezed on both arms after the beamsplitter.  

Crucially both the variances and the correlations measured in the Gaussian EB scheme correspond exactly to measurable quantities in the P\&M version.   Having obtained these quantities , namely Alice and Bob each measuring $\hat{x}$ and $\hat{p}$, we wish to infer back through the operations at their stations to completely characterise the two-mode state shared by Alice and Bob directly after transmission and hence quantify Eve's information.  Bob and Alice's operations are just beamsplitters and single mode-squeezers so this can be straightforwardly achieved at the level of the covariance matrix using appropriate symplectic transformations.  Note that Eve's information evaluated with respect to the four mode state just prior to measurement will be exactly the same as that of the two mode state as the evolution between the two points involves only pure states and unitary operations.  The covariance matrices before and after Alice and Bob's beamsplitters are related by,
\eq{\nn\gamma_f = {\bf S}^T(r){\bf BS}^T\bk{T_B}{\bf BS}^T\bk{T_A}\gamma_i{\bf BS}\bk{T_A}{\bf BS}\bk{T_B}{\bf S}(r)}
where 
\eqn{\nn {\bf S}(r) = \mat{e^{-r}}{0}{0}{e^r}, \hs {\bf BS}(T) = \mat{\sqrt{T}\mathbb{I}_2}{\sqrt{1-T}\mathbb{I}_2}{-\sqrt{1-T}\mathbb{I}_2}{\sqrt{T}\mathbb{I}_2}}
are the symplectic transformations for a one-mode squeezer and a two-mode beamsplitter respectively.  Using our knowledge of Alice and Bob's station, namely that $T_A = T_B = \half$ and that in $\gamma_i$ the ancillae are uncorrelated vacua we can deduce an equivalent 2-mode covariance matrix in the standard form of an EPR with one arm distributed though a Gaussian channel,
\eqn{\gamma_{AB}\label{sym} = \mat{1+ V_\alpha\hs\mathbb{I}_2}{\sqrt{\eta(V_\alpha^2+2V_\alpha)} \hs \sigma_z}{\sqrt{\eta(V_\alpha^2+2V_\alpha)}\hs\sigma_z}{\hs \eta V_\alpha+ \eta\delta +1\hs\mathbb{I}_2}}
where the parameters $\eta$,$\delta$ and $V_\alpha$ characterise an effective Gaussian ensemble sent though a Gaussian channel.  The crucial tradeoff in this scheme is between a large post-selection to improve the effective channel parameters and the probability of obtaining measurement results within the post-selection region.  Note that in order to safely assume that the covariance matrix is of the symmetric form, Eq.\ref{sym}, Alice and Bob will need to make use of an active symmetrisation step as outlined in \cite{Leverrier:2011p1357}.

In particular although the post-selection is symmetric about the chosen quadratures in phase space it is not generally rotationally symmetric.  In \cite{Leverrier:2011p1357} the authors solve this issue which similarly effects discretely modulated protocols by having Alice and Bob implement a series of randomly chosen orthogonal transformations on their classical data.  This is equivalent to performing random conjugate transformations in the EB scheme and allows Alice and Bob to essentially enforce the symmetry they want.  In the P\&M version of our protocol these transformations are performed after measurement but before the post-selection.  This means that although Eve is assumed to know the absolute values of the post-selection threshold she does not know the direction in phase space upon which they will be enforced, preventing her from gaining an advantage through symmetry breaking.
 
In general the result of the post-selection is to produce a much larger initial EPR state that has gone through better effective channel.  This occurs becuase we are post-selecting on the data for which Alice and Bob are more correlated.  Due to the non-deterministic nature of the protocol the information that could be extracted from the whole ensemble is sharply decreased but the information advantage on the post-selected key is higher.

{\it Gaussian channel.---}
Having established a method of evaluating security under general eavesdropper attacks it remains to see what kind of performance this method results in for the most common case, the noisy Gaussian channel. Such a channel is completely parameterised by it's transmission $T$ and excess thermal noise $\xi$.  In this case there is a class of collective eavesdropper attacks known to be equivalent and optimal and we are free to choose a specific form to facilitate calculation \cite{Pirandola:2008p1765}.  We choose the entangling cloner attack where we assume Eve replaces the lossy channel with a perfect channel and a beamsplitter of transmission $T$ on which she mixes Alice's incoming signal with one arm of her own EPR pair.  Both EPR states are generated by combining finitely $\hat{x}$ and $\hat{p}$-squeezed states on balanced beamsplitters. The squeezing of Eve's state is determined by the channel according to $V_E = \exp\bk{\cosh^{-1}(1-T+\xi)/(1-T)}$ while the squeezing of Alice's EPR is fixed by the variance of her classical distribution according to $V_S = \exp\bk{\cosh^{-1}(V_A+1)}$.  We now evolve the 6-mode (4 EPR arms and 2 vacuum ancillae) system through channel, beamsplitters and post-selection to obtain that state just before the homodyne detectors,
\begin{widetext}
\eqn{\nn&\ket{ABE}& = \int dx_i \hs \psi(x_i)\ket{\frac{x_1 - x_2}{2} + \frac{x_3}{\sqrt{2}}}_A\ket{\frac{x_1 - x_2}{2} - \frac{x_3}{\sqrt{2}}}_A \ket{\sqrt{\frac{T}{2}}(x_4+x_5) + \sqrt{\frac{1-T}{2}}(x_1+x_2)}_E\ket{\frac{x_4-x_5}{\sqrt{2}}}_E\\
& & \ket{\frac{\sqrt{T}}{2}(x_1+x_2) - \frac{\sqrt{1-T}}{2}(x_4+x_5) + \frac{x_6}{\sqrt{2}}}_B
\ket{\frac{\sqrt{T}}{2}(x_1+x_2) - \frac{\sqrt{1-T}}{2}(x_4+x_5) - \frac{x_6}{\sqrt{2}}}_B, \hs i=1...6}
\end{widetext}
where \eq{\psi(x_i) = \frac{1}{(2\pi)^{3/2}}e^{\bk{-\frac{x_1^2}{4V_S} - \frac{x_2^2V_S}{4} - \frac{x_3^2}{4} - \frac{x_4^2}{4V_E} - \frac{x_5^2V_E}{4} - \frac{x_6^2}{4}}}} and the subscripts on the kets indicate ownership.  The post-selection involves taking definite integrals over Alice and Bob's modes and the covariance matrix elements corresponding to the measured quadratures are calculated according to $\gamma_{ij} = \half\var{\{\hat{x}_i\hat{x}_j \}}-\var{\hat{x}_i}\var{\hat{x}_j}$ over $\ket{ABE}$.  Having calculated these quantities we calculate security for the symmetric Gaussian state with the same covariance matrix as outlined above.  The final step is to multiply this keyrate by the probability of the post-selection succeeding, which in the EB scheme is simply the norm of the post-selected state.

\begin{figure}[htbp]
\begin{center}
\includegraphics[width = 8cm]{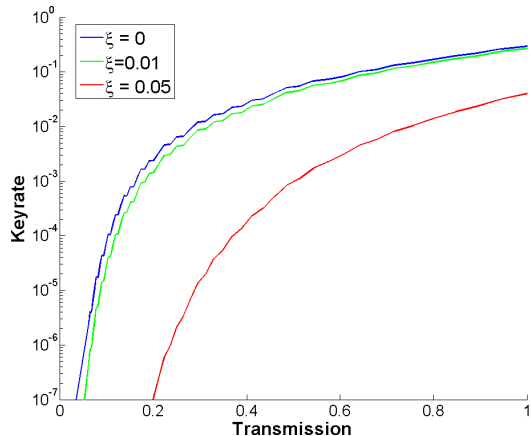}
\caption{\label{K} Secret key rates for direct reconciliation combined with post-selection as a function of transmission for increasing levels of excess noise ($\xi = 0$, $\xi=0.01$, $\xi = 0.05$) with decreasing keyrate.}
\end{center}
\end{figure}

Results for keyrate as a function of channel transmission are given in Fig.\ref{K} for realistic levels of thermal noise on the channel, where for each pair of channel parameters the post-selection is numerically optimised over.  Encouragingly we see that noise on the order of 1\% results in only a modest reduction of the keyrate.

{\it Conclusions.---}

In conclusion we have shown that post-selection based CVQKD is secure for arbitrary collective attacks, and thus asymptotically secure for all attacks.  This was achieved by identifying an entanglement based scheme that correctly reflects the post-selected ensemble that is used in the final key generation.  Furthermore these results show little loss of performance compared to previous analyses of post-selection and are quite robust to modest amounts of excess noise on the channel.  There are several avenues for further work, foremost being the fact that the bounds given here are certainly not tight.  Other topics of interest would be the incorporation of finite-size effects and the combination of post-selection with other protocols to identify the optimal technique for a given scenario.

The authors would like to thank Norbert L\"utkenhaus, Christian Weedbrook and Anthony Leverrier for helpful discussions.  This research was conducted by the Australian Research Council Centre of Excellence for Quantum Computation and Communication Technology (Project number CE11000102

\bibliographystyle{apsrev}
\bibliography{paper}
\end{document}